\documentclass{aa}

\usepackage{graphicx}
\usepackage{astron}

\begin{document}

\thesaurus{11(09.11.1;09.19.1;11.08.1;11.11.1;11.19.6)}

\title{Continuous Fields and Discrete Samples:\\
Reconstruction through Delaunay Tessellations}
\titlerunning{Delaunay Reconstruction of Continuous Fields}

\author{W.E.~Schaap \and R.~van~de Weygaert}

\institute{Kapteyn Astronomical Institute, University of Groningen,
  P.O.~Box 800, 9700 AV, Groningen, The Netherlands}


\date{Received ~~~~~~~~~~~~~~~~~~~~~~~ / Accepted }

\maketitle

\begin{abstract}
  Here we introduce the Delaunay Density Estimator Method. Its
  purpose is rendering a fully volume-covering reconstruction of a
  density field from a set of discrete data points sampling this
  field. Reconstructing density or intensity fields from a set of
  irregularly sampled data is a recurring key issue in operations on
  astronomical data sets, both in an observational context as well as
  in the context of numerical simulations. Our technique is based upon
  the stochastic geometric concept of the Delaunay tessellation
  generated by the point set. We shortly describe the method, and
  illustrate its virtues by means of an application to an N-body
  simulation of cosmic structure formation. The presented technique is
  a fully adaptive method: automatically it probes high density
  regions at maximum possible resolution, while low density regions
  are recovered as moderately varying regions devoid of the often
  irritating shot-noise effects.  Of equal importance is its
  capability to sharply and undilutedly recover anisotropic density
  features like filaments and walls.  The prominence of such features
  at a range of resolution levels within a hierarchical clustering
  scenario as the example of the standard CDM scenario is shown to be
  impressively recovered by our scheme.

\end{abstract}

\vspace*{-.2cm}

\section{Introduction}
Astronomical observations, physical experiments as well as computer
simulations often involve discrete data sets supposed to represent a
fair sample of an underlying smooth and continuous field. 
Conventional methods are usually plagued by one or more 
artefacts. Firstly, they often involve estimates at a restricted 
and discrete set of locations -- usually defined by a grid -- 
instead of a full volume-covering field reconstruction. A 
problem of a more fundamental nature is that the resulting 
estimates are implicitly mass-weighted averages, whose 
comparison with often volume-weighted analytical quantities 
is far from trivial. For most practical purposes, the 
disadvantage of almost all conventional methods is their 
insensitivity and inflexibility to the sampling point process.  
This leads to a far from optimal performance in both high density 
and low density regions, which often is dealt with by rather 
artificial and ad hoc means. 

In particular in situations of highly non-uniform distributions 
conventional methods tend to conceal various interesting and 
relevant aspects present in the data. The cosmic matter 
distribution exhibits conspicuous features like filaments 
and walls, extended along one or two directions while 
compact in the other(s). In addition, the density fields 
display structure of varying contrasts over a large range 
of scales. Ideally sampled by the data points, appropriate 
field reconstructions should be set solely and automatically 
by the point distribution itself. The commonly used methods, 
involving artificial filtering through for instance grid size 
or other smoothing kernels (e.g. Gaussian filter)
often fail to achieve an optimal result. 

\begin{figure*}
\begin{picture}(100,150)(0,10)
  \includegraphics[width=12cm,keepaspectratio]{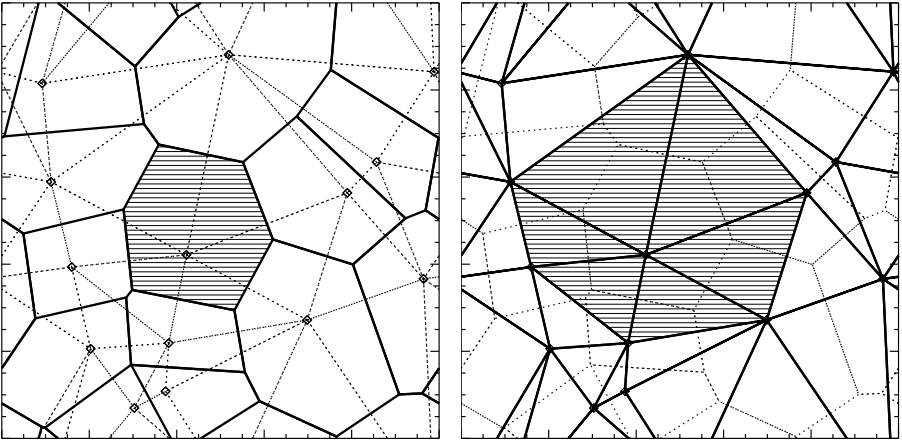}
\end{picture}
\hfill \parbox[b]{5.5cm}{ \caption{\small A set of 20 points with
    their Voronoi (left frame: solid lines) and Delaunay (right frame:
    solid lines) tesselations. Left frame: the shaded region indicates
    the Voronoi cell corresponding to the point located just below the
    center. Right frame: the shaded region is the ``contiguous Voronoi
    cell'' of the same point as in the lefthand frame.
    }\vspace*{-.3cm} }
\end{figure*}

Here we describe and propagate a new fully self-adaptive method 
based on the Delaunay triangulation of the given point process. 
After a short description of the fundamentals of our tessellation 
procedure, we show its convincing performance on the result 
of an N-body simulation of structure formation, whose particle 
distribution is supposed to reflect the underlying cosmic 
density field. A detailed specification of the method, together 
with an extensive quantitative and statistical evaluation of 
its performance will be presented in a forthcoming publication 
\cite{schetwey20}. 

\section{The Delaunay Tessellation Field Estimator}
Given a set of field values sampled at a discrete number of 
locations along one dimension we are familiar with various 
prescriptions for reconstructing the field over the full spatial 
domain. The most straightforward way involves the partition of 
space into bins centered on the sampling points. The field is then  
assumed to have the -- constant -- value equal to the one at the 
sampling point. Evidently, this yields a field with unphysical 
discontinuities at the boundaries of the bins. A first-order 
improvement concerns the linear interpolation between the sampling 
points, leading to a fully continuous field. 

In more than one dimension, the equivalent spatial intervals of the
1-D bins are well-known in stochastic geometry. A point process
defines a Voronoi tessellation by dividing space into a unique and
volume-covering network of mutually disjunct convex polyhedral cells,
each of which comprises that part of multidimensional space closer to
the defining point than to any of the other (see van de Weygaert
\cite*{wey91} and references therein). These Voronoi cells (see
Fig.~1) are the multidimensional generalization of the 1-D bins in
which the zeroth-order method approximates the field value to be
constant.  The natural extension to a multidimensional linear
interpolation interval then immediately implies the corresponding
Delaunay tessellation \cite{del34}. This tessellation (Fig.~1)
consists of a volume-covering tiling of space into tetrahedra (in 3-D,
triangles in 2-D, etc.) whose vertices are formed by four specific
points in the dataset. The four points are uniquely selected such that
their circumscribing sphere does not contain any of the other
datapoints. The Voronoi and Delaunay tessellation are intimately
related, being each others dual in that the centre of each Delaunay
tetrahedron's circumsphere is a vertex of the Voronoi cells of each of
the four defining points, and conversely each Voronoi cell nucleus a
Delaunay vertex (see Fig.~1). The ``minimum triangulation'' property
of the Delaunay tessellation has in fact been well-known and
abundantly applied in, amongst others, surface rendering applications
such as geographical mapping and various computer imaging algorithms.

Consider a set of ${\rm N}$ discrete datapoints in a finite region of 
M-dimensional space. Having at one's disposal the field values at each of 
the $(1\!+\!{\rm M})$ Delaunay vertices ${\bf x}_0, {\bf x}_1, \ldots, 
{\bf x}_{\rm M}$, at each 
location ${\bf x}$ in the interior of a Delaunay M-dimensional 
tetrahedron the linear interpolation field value is defined by 
\begin{equation}
  f({\bf x})~=~f({\bf x}_0)~+~ \left. \nabla f ({\bf x}_0)\right|_{\rm Del} \!\!\!\cdot ({\bf x}- {\bf x}_0)\,,
\end{equation}
in which $\left. \nabla f ({\bf x}_0) \right|_{\rm Del}$ 
is the estimated constant field gradient within the tetrahedron. 
Given the $(1\!+\!{\rm M})$ field values  $f({\bf x}_0), f({\bf x}_1), \ldots, 
f({\bf x}_{\rm M})$, the value of the ${\rm M}$ components of $\left. \nabla f ({\bf x}_0) \right|_{\rm Del} $ can be computed straightforwardly 
by evaluating Eqn.~(1) for each of the ${\rm M}$ points 
${\bf x}_1, \ldots, {\bf x}_{\rm M}$. This multidimensional procedure of 
linear interpolation was already described by Bernardeau \& van de 
Weygaert \cite*{beretwey96} in the context of defining procedures 
for volume-weighted estimates of cosmic velocity fields. While they 
explicitly demonstrated that the zeroth-order Voronoi estimator 
is the asymptotic limit for volume-weighted field reconstructions 
from discretely sampled field values, they showed the superior 
performance of the first-order Delaunay estimator in reproducing 
analytical predictions. 

The one factor complicating a trivial and direct implementation of
above procedure in the case of density (intensity) field estimates is
the fact that the number density of data points itself is the measure
of the underlying density field value. Unlike the case of velocity
fields, we therefore cannot start with directly available field
estimates at each datapoint. Instead, we need to define appropriate
estimates from the point set itself.  Most suggestive would be to base
the estimate of the density field at the location ${\bf x}_i$ of each
point on the inverse of the volume ${\rm V_{Vor,i}}$ of its Voronoi
cell, $\rho({\bf x}_i)\!=\!{\rm m}/{\rm V_{Vor,i}}$. Note that in this
we take every datapoint to represent an equal amount of mass ${\rm m}$.
The resulting field estimates are then intended as input for the above
Delaunay interpolation procedure. However, one can demonstrate that
integration over the resulting density field would yield a different
mass than the one represented by the set of sample points (see Schaap
\& van de Weygaert \cite*{schetwey20} for a more specific and detailed
discussion).  Instead, mass conservation is naturally guaranteed when
the density estimate is based on the inverse of the volume ${\rm
  W_{Vor,i}}$ of the ``contiguous'' Voronoi cell of each datapoint,
$\rho({\bf x}_i)\!\propto\!1/{\rm W_{Vor,i}}$. The ``contiguous''
Voronoi cell of a point is the cell consisting of the agglomerate of
all $K$ Delaunay tetrahedra containing point $i$ as one of its
vertices, whose volume ${\rm W_{Vor,i}}=\sum_{j=1}^K {\rm V_{Del,j}}$
is the sum of the volumes ${\rm V_{Del,j}}$ of each of the $K$
Delaunay tetrahedra. Figure~1 (righthand panel) depicts an
illustration of such a cell. Properly normalizing the mass contained
in the reconstructed density field, taking into account the fact that
each Delaunay tetrahedron is invoked in the density estimate at
$1\!+\!{\rm M}$ locations, we find at each datapoint the following
density estimate,
\begin{equation}
  \rho({\bf x}_i)~=~{\rm {m\,(1\!+\!M)}}/{{\rm W_{Vor,i}}}
\end{equation}
Having computed these density estimates, we subsequently proceed to
determine the complete volume-covering density field reconstruction
through the linear interpolation procedure outlined in Eqn.~(1).

\section{Analysis of a cosmological N-body simulation}
Cosmological N-body simulations provide an ideal template for
illustrating the virtues of our method. They tend to contain a large
variety of objects, with diverse morphologies, a large reach of
densities, spanning over a vast range of scales. They display low
density regions, sparsely filled with particles, as well as highly
dense and compact clumps, represented by a large number of particles.
Moderate density regions typically include strongly anisotropic
structures such as filaments and walls.

\begin{figure*}
\begin{picture}(-50,380)(0,0)

\end{picture}
\hfill \parbox[b]{4.cm}{ \caption{\small A 9-frame mosaic comparing
    the performance of the Delaunay density estimating technique with
    a conventional grid-based TSC method in analyzing a cosmological
    N-body simulation. Left column: the particle distribution in a
    $10h^{-1}\hbox{Mpc}$ wide central slice through the simulation
    box. Central column: the corresponding Delaunay density field
    reconstruction. Right column: the TSC rendered density field
    reconstruction. The colour scale of the density fields is
    logarithmic, ranging from $\delta\rho/\rho=0-2400$.}\vspace*{.5cm}
  }
\vspace*{-.2cm}
\end{figure*}

Each of these features have their own individual characteristics, and
often these may only be sufficiently highlighted by some specifically
designed analysis tool. Conventional methods are usually only tuned
for uncovering one or a few aspects of the full array of properties.
Instead of artificial tailor-made methods, which may be insensitive to
unsuspected but intrinsically important structural elements, our
Delaunay method is uniquely defined and fully self-adaptive. Its
outstanding performance is clearly illustrated by Figure~2. Here we
have analyzed an N-body simulation of structure formation in a
standard CDM scenario ($\Omega_0$\,=\,1, $H_0$\,=\,50 km/s/Mpc). It
shows the resulting distribution of $128^3$ particles in a cubic
simulation volume of 100$h^{-1}$ Mpc, at a cosmic epoch at which
$\sigma(R_{TH}=8h^{-1} \mbox{Mpc})=1$. The figure depicts a
$10h^{-1}\hbox{Mpc}$ slice through the center of the box. The
lefthand column shows the particle distribution in a sequence of frames at
increasingly fine resolution. Specifically we zoomed in on the
richest cluster in the region. The righthand column shows the
corresponding density field reconstruction on the basis of the
grid-based Triangular-Shaped Clouds (TSC) method, here evaluated on a
$518^2$ grid. For the TSC method, one of the most frequently applied
algorithms, we refer to the description in Hockney and Eastwood
\cite*{hoceteas81}. A comparison with other, more elaborate methods
which have been developed to deal with the various aspects that we
mentioned, of which Adaptive Grid methods and SPH based methods have
already acquired some standing, will be presented in Schaap and van de
Weygaert \cite*{schetwey20}.

A comparison of the lefthand and righthand columns with the central
column, i.e.~the Delaunay estimated density fields, reveals the
striking improvement rendered by our new procedure. Going down from
the top to the bottom in the central column, we observe seemingly
comparable levels of resolved detail. The self-adaptive skills of the
Delaunay reconstruction evidently succeed in outlining the full
hierarchy of structure present in the particle distribution, at every
spatial scale represented in the simulation. The contrast with the
achievements of the fixed grid TSC method in the righthand column is
striking, in particular when focus tunes in on the finer structures.
The central cluster appears to be a mere featureless blob! In
addition, low density regions are rendered as slowly varying regions
at moderately low values. This realistic conduct should be set off
against the erratic behaviour of the TSC reconstructions, plagued by
annoying shot-noise effects.

Figure~2 also bears witness to another virtue of the Delaunay
technique. It evidently succeeds in reproducing sharp, edgy and clumpy
filamentary and wall-like features. Automatically it resolves the fine
details of their anisotropic geometry, seemlessly coupling sharp
contrasts along one or two compact directions with the mildy varying
density values along the extended direction(s). Moreover, it also
manages to deal succesfully with the substructures residing within
these structures. The well-known poor operation of e.g.~the TSC method
is clearly borne out by the central righthand frame. Its fixed and
inflexible ``filtering'' characteristics tend to blur the finer
aspects of such anisotropic structures. Such methods are therefore
unsuited for an objective and unbiased scrutiny of the foamlike
geometry which so pre-eminently figures in both the observed galaxy
distribution as well as in the matter distribution in most viable
models of structure formation.

Not only qualitatively, but also quantitatively our method turns out
to compare favourably with respect to conventional methods. We are in
the process of carefully scrutinizing our method by means of an array
of quantitative tests. A full discussion will be presented in Schaap
and van de Weygaert \cite*{schetwey20}. Here we mention the fact that
the method recovers the density distribution function over many orders
of magnitude. The grid-based methods, on the other hand, only managed
to approach the appropriate distribution in an asymptotic fashion and
yielded reliable estimates of the distribution function over a mere
restricted range of density values. Very importantly, on the basis of
the continuous density field reconstruction of our Delaunay method, we
obtained an estimate of the density autocorrelation function that
closely adheres to the (discrete) two-point correlation function
directly determined from the point distribution. Further assessments
on the basis of well-known measures like the Kullback-Leibler
divergence \cite{kuletlei51}, an objective statistic for quantifying
the difference between two continuous fields, will also be presented
in Schaap and van de Weygaert \cite*{schetwey20}. Finally, we may also
note that in addition to its statistical accomplishments, we should
also consider the computational requirements of the various methods.
Given a particle distribution, the basic action of computing the
corresponding Delaunay tessellation, itself an ${\mathcal O}(N)$
routine \cite{wey91}, the subsequent interpolation steps, at any
desired resolution, are considerably less CPU intensive than the TSC
method (both also ${\mathcal O}(N)$). In the case of figure 2 the
Delaunay method is about a factor of 10 faster. In the present
implementation, the bottleneck is Delaunay's substantial memory
requirement ($\approx 10 \times $ the TSC operation), but a more
efficient algorithm will be available in short order.   These issues
will be treated extensively in our upcoming publication.

The preceding is ample testimony of the promise of tessellation
methods for the aim of continuous field reconstruction. The presented
method, following up on earlier work by Bernardeau \& van de Weygaert
\cite*{beretwey96}, may be seen as a first step towards yet more
advanced tessellation methods. One suggested improvement will be a
second-order method rendering a continuously differentiable field
reconstruction, which would dispose of the rather conspicuous
triangular patches that form an inherent property of the linear
procedure with discontinous gradients. In particular, we may refer to
similar attempts to deal with related problems, along the lines of
natural neighbour interpolation \cite{sib81}, such as implemented in
the field of geophysics \cite{sametal95,braetsam95} and in engineering
mechanics \cite{suk98}. As multidimensional discrete data sets are a
major source of astrophysical information, we wish to promote such
tessellation methods as a natural instrument for astronomical data
analysis.

\begin{acknowledgements}
We thank E.~Romano-D\'{\i}az for substantial contributions,
F.~Bernardeau for instigating this line of research, and V.~Icke and
B.~Jones for useful discussions.
\end{acknowledgements}

\bibliographystyle{astron}
\bibliography{mnemonic,references}
\end{document}